\documentclass[prd,twocolumn,aps,showpacs,nofootinbib,nobibnotes,superscriptaddress,preprintnumbers]{revtex4}
\usepackage{epsfig}
\usepackage{graphics}
\usepackage{bm}
\usepackage{color}
\usepackage{dcolumn}
\usepackage{bm}
\usepackage{bbm}
\usepackage{amssymb}
\usepackage{amsmath}
\usepackage{latexsym}
\usepackage{float}
\usepackage{ifthen}
\usepackage{caption,subfig}
\usepackage{enumerate}
\usepackage{url}
\usepackage{caption,subfig}
\usepackage{amsopn}
\usepackage{hyperref}

\usepackage{amsfonts}
\usepackage{multirow}
\usepackage{array}
\usepackage{booktabs}
\usepackage{rotating}

\usepackage{ulem}
\def\clap#1{\hbox to 0pt{\hss#1\hss}}

\def\({\left(}
\def\){\right)}
\def\[{\left[}
\def\]{\right]}
\def\bea{\begin{eqnarray}}
\def\eea{\end{eqnarray}}
\def\be{\begin{equation}}
\def\ee{\end{equation}}
\def\ba{\begin{eqnarray}}
\def\ea{\end{eqnarray}}
\def\beq{\begin{eqnarray}}
\def\eeq{\end{eqnarray}}

\def\d{\mathrm{d}}
\newcommand{\cs}{c_s}

\newcommand{\mM}{\mathcal{M}}
\newcommand{\mH}{\mathcal{H}}
\newcommand{\Ft}{\tilde{F}}

\renewcommand{\geq}{\geqslant}

\def\p{\partial}

\def\cs{c_{\rm s}}

\def\be{\begin{equation}}
\def\ee{\end{equation}}
\def\ba{\begin{eqnarray}}
\def\ea{\end{eqnarray}}
\def\beq{\begin{eqnarray}}
\def\eeq{\end{eqnarray}}

\def\d{\mathrm{d}}
\def\p{{\cal P}}

\def\L*{{\cal L}_*}
\def\L{\mathcal{L}}
\def\({\left(}
\def\){\right)}

\def\nn{\nonumber}
\def\p{\partial}
\def\mn{_{\mu \nu}}

\def\stu{St\"uckelberg }
\def\p{\partial}

\def\<{\langle}
\def\>{\rangle}

\def\hA{{\hat A}}

\def\cs2{c_{s}^{2}}

 \def\p{\partial}

\def\be{\begin{equation}}
\def\ee{\end{equation}}
\def\ba{\begin{eqnarray}}
\def\ea{\end{eqnarray}}
\def\beq{\begin{eqnarray}}
\def\eeq{\end{eqnarray}}

\def\d{\mathrm{d}}
\def\p{{\cal P}}

\def\L*{{\cal L}_*}
\def\L{\mathcal{L}}
\def\({\left(}
\def\){\right)}

\def\nn{\nonumber}
\def\p{\partial}
\def\mn{_{\mu \nu}}
\def\stu{St\"uckelberg }
\def\p{\partial}

\def\<{\langle}
\def\>{\rangle}

\newcommand{\Lag}{{\mathcal{L}}}

\definecolor{hyperref}{RGB}{026,028,087}

\begin{document}

\title{Generalized Proca \& its Constraint Algebra

}

\date{\today,~ $ $}

\author{Jose Beltr\'an Jim\'enez} \email{jose.beltran@usal.es}
\affiliation{Departamento de F\'isica Fundamental and IUFFyM, Universidad de Salamanca, E-37008 Salamanca, Spain.}

\author{Claudia de Rham} \email{c.de-rham@imperial.ac.uk}
\affiliation{Theoretical Physics, Blackett Laboratory, Imperial College, London, SW7 2AZ, U.K.}
\affiliation{CERCA, Department of Physics, Case Western Reserve University, 10900 Euclid Ave, Cleveland, OH 44106, USA}

\author{Lavinia Heisenberg} \email{lavinia.heisenberg@phys.ethz.ch}
\affiliation{Institute for Theoretical Physics,
ETH Zurich, Wolfgang-Pauli-Strasse 27, 8093, Zurich, Switzerland}

\date{\today}

\begin{abstract}
We reconsider the construction of general derivative self-interactions for a massive Proca field. The constructed Lagrangian is such that the vector field propagates at most three degrees of freedom, thus avoiding the ghostly nature of a fourth polarisation. The construction makes use of the well-known condition for constrained systems of having a degenerate Hessian. We briefly discuss the casuistry according to the nature of the existing constraints algebra. We also
explore various classes of interesting new interactions that have been recently raised in the literature. For the sixth order Lagrangian that satisfies the constraints by itself we prove its topological character, making such a term irrelevant. There is however a window of opportunity for exploring other classes of fully-nonlinear interactions that satisfy the constraint algebra by mixing terms of various order.
\end{abstract}


\maketitle

\section{Introduction}
The physical mechanism behind the cosmic acceleration is still under investigation. It could be due to a cosmological constant or a time-varying additional degree of freedom. In the latter case, much effort has been put in constructing viable infrared modifications of gravity \cite{deRham:2014zqa,Heisenberg:2018vsk}. The simplest scenarios consist of an additional scalar field as part of the modification, where the scalar field does not need necessarily to be a canonical scalar field with a potential but actually can also have second order derivative interactions, as it is the case for the class of Galileon theories \cite{Nicolis:2008in}. In order to avoid related Ostrogradski instabilities, the resulting field equations must propagate at most one additional scalar mode. The constructed interactions are invariant under a constant shift of the field and its gradient. These symmetries are abandoned as soon as the interactions are generalized to the curved space-time case, resorting to Horndeski interactions \cite{Horndeski:1974wa}. They constitute the most general action for a scalar-tensor theory with second order equations of motion. A nice property of the Galileon and Horndeski interactions is that they can be constructed using the symmetry properties of the Levi-Civita tensor, very much like the Lovelock terms for gravity.

The attempt to construct Galilean interactions for a massless spin-1 field met immediately a no-go theorem \cite{Deffayet:2010zh}. Nevertheless, this finding can be avoided if one is willing to give up the gauge invariance of the vector field. In this way it became possible to construct derivative vector self-interactions while still keeping three propagating degrees of freedom. This opens up a new avenue for phenomenological applications in cosmology, astrophysics and black hole physics \cite{phenomeno}.
 In \cite{Jimenez:2016isa} the general Lagrangian for a massive vector field with derivative self-interactions and propagating the required three polarisations was systematically constructed. The consistency on these interactions was shown in two different approaches, which also confirmed in a complementary way their completeness. The constructed interactions were firstly analyzed in a bottom-up fashion using the decoupling limit and explicitly requiring the necessary conditions for the transverse and longitudinal modes to satisfy in order to guarantee the right number of propagating degrees of freedom. It was specially shown that the pure \stu field sector (the longitudinal mode) belongs to the Galileon class of Lagrangians with shift symmetry and the mixed couplings between the transverse and longitudinal modes can only happen via specific highly constrained symmetric rank--2 and rank--4 tensors.

 These findings were confirmed and complemented by the analysis of a systematic construction beyond the decoupling limit using the antisymmetric properties of the Levi-Civita tensor \cite{Jimenez:2016isa}. Order by order in derivatives of the vector field, the complete set of allowed interactions was successfully presented. In this note we will discuss the possibility of having a degenerate Hessian guaranteeing the propagation of at most three polarisations, not order by order in derivatives of the field, but through non-trivial cancellations among different orders. We comment on the possibility of relating such constructed interactions to the standard ones by means of field redefinitions, which becomes relevant after setting the coupling to external sources. We also discuss the role of duality relations in these comparisons.

  \section{Generalized Proca}
Introducing the first derivative of the vector field as $B_{\mu\nu}=\partial_\mu A_\nu$, we can build the antisymmetric $F_{\mu\nu}=B_{\mu\nu}-B_{\nu\mu}$ field strength and the symmetric derivative tensor $S_{\mu\nu}=B_{\mu\nu}+B_{\nu\mu}$. In \cite{Jimenez:2016isa}, it was shown that the only allowed operators for the mixing of the transverse and longitudinal modes are $\tilde{F}^{\mu\alpha}\tilde{F}_\alpha^\nu S_{\mu\nu}$ and $\tilde{F}^{\mu\nu}\tilde{F}^{\alpha\beta}S_{\mu\alpha}S_{\nu\beta}$ (note $S_{\mu\nu}\to\partial_{\mu}\partial_\nu\pi$ in the decoupling limit). In this way the second time derivatives of the longitudinal mode couple only to the magnetic part of the gauge field and higher order equations of motion are avoided.

The decoupling limit analysis and the systematic construction order by order in $B_{\mu\nu}$ yielded the following general Lagrangian \cite{Heisenberg:2014rta,Jimenez:2016isa}:
 \begin{equation}\label{generalizedProfaField}
\mathcal L_{\rm gen. Proca} =\sum^5_{n=2}\alpha_n \mathcal L_n \,,
\end{equation}
where the self-interactions of the vector field are
\begin{eqnarray}\label{vecGalProcaField}
\mathcal L_2 & = &f_2(A_\mu, F_{\mu\nu}, \tilde{F}_{\mu\nu})\nonumber\\
\mathcal L_3 & = &f_3(A^2) \; [B] \nonumber\\
\mathcal L_4  &=&  f _4(A^2)\;\left([B]^2-[B^2]\right)   \nonumber\\
\mathcal L_5  &=&f_5(A^2)\;\left([B]^3-3[B][B^2]+2[B^3] \right] \nonumber\\
&& +\tilde{f}_5(A^2)\tilde{F}^{\alpha\mu}\tilde{F}^\beta_{\;\;\mu} B_{\alpha \beta} \nonumber\\
\mathcal L_6  &=&f_6(A^2) \tilde{F}^{\alpha\beta}\tilde{F}^{\mu\nu}B_{\alpha \mu}B_{\beta \nu} \,,
\end{eqnarray}
where square brackets designate the trace of a tensor and  $\tilde{F}={}^*\!F$ is the Hodge-dual of the Maxwell tensor. The breaking of parity in the above Lagrangians can only occur in $\Lag_2$ \cite{Jimenez:2016isa,Allys:2016jaq}. The higher order Lagrangians are even functions of $\tilde{F}$ so they do not break parity.
Since the antisymmetric part of $B_{\mu\nu}$ in the above interactions will only contribute terms that can be all included in $\mathcal L_2$, we can replace $B_{\mu\nu}$ by $S_{\mu\nu}$ in equation \eqref{vecGalProcaField}.

 The sixth order Lagrangian deserves a special mention here. It was first omitted in \cite{Heisenberg:2014rta}, where the condition was followed that the longitudinal mode should not have any trivial total derivative. Relaxing this condition allowed the construction of the sixth order Lagrangian of generalized Proca in \cite{Allys:2015sht}, which was also confirmed in \cite{Jimenez:2016isa}.

Following the construction scheme of \cite{Jimenez:2016isa}, at this order in interactions, there are two possible contractions
\begin{align}\label{LagrangianL6}
\mathcal{L}_6=-\mathcal{L}_6^T-e_2\mathcal{L}_6^C,
\end{align}
where
\begin{align}\label{LagrangianL6ind}
\mathcal{L}_6^T=&f_6(A^2)\epsilon^{\mu\nu\rho\sigma}\epsilon^{\alpha\beta\delta\kappa}B_{\mu \alpha}B_{\nu \beta}B_{\rho \delta} B_{\sigma \kappa} \nonumber\\
\mathcal{L}_6^C=&\tilde{f}_6(A^2)\epsilon^{\mu\nu\rho\sigma}\epsilon^{\alpha\beta\delta\kappa}B_{\mu \nu}B_{\alpha \beta} B_{\rho \delta}B_{\sigma \kappa}\,.
\end{align}
While the first contraction gives rise to contributions in form of pure $S$, or pure $F$, or mixed $\tilde{F}\tilde{F}SS$, the second contraction generates only the latter two. In \cite{Jimenez:2016isa}, their contributions were shown explicitly to be
\begin{align}\label{eqL6Tausgeschrieben}
\mathcal{L}_6^T &=\frac{-f_6}{16}\Big(3([F^2]^2-2[F^4])-12\Ft^{\mu\nu}\Ft^{\alpha\beta}S_{\mu\alpha}S_{\nu\beta} \nonumber\\
& +[S]^4 -6[S]^2[S^2]+3[S^2]^2+8[S][S^3]-6[S^4]\Big)
\end{align}
for the first contraction and
\ba\label{eqL6Causgeschrieben}
\mathcal{L}_6^C  =\frac{-\tilde{f}_6}{16}\Big(2([F^2]^2-2[F^4])-4\Ft^{\mu\nu}\Ft^{\alpha\beta}S_{\mu\alpha}S_{\nu\beta}\Big)
\ea
for the second contraction.
In fact, there is also the contraction $\epsilon^{\mu\nu\rho\sigma}\epsilon^{\alpha\beta\delta\kappa}B_{\mu \nu}B_{\alpha \beta}B_{\rho \sigma}B_{\delta \kappa}$ but this has been discarded in \cite{Jimenez:2016isa} since it gives rise to purely gauge invariant quantities, already included in $\mathcal{L}_2$.

The expression \eqref{eqL6Tausgeschrieben} was considered at various places in the literature and has received a particular recent interest. First introduced in \cite{Allys:2015sht,Jimenez:2016isa} (see also \cite{Heisenberg:2018vsk}), it was immediately discarded for being irrelevant in four-dimensions. Indeed, based on decoupling limit reasoning,  it is clear that this term could never lead to any genuinely new interaction as was argued in \cite{Jimenez:2016isa}. Yet motivated by the interesting phenomenology and implications of Proca interactions \cite{phenomeno}, such a term was recently resurrected in \cite{ErrastiDiez:2019ttn} (in four-dimensional flat spacetime). For completeness and to prevent any further confusion we re-consider this term \eqref{eqL6Tausgeschrieben} in what follows and explicitly show that {\it in four-dimensional flat space-time} this term is indeed a total derivative and hence irrelevant as previously argued in \cite{Jimenez:2016isa,Heisenberg:2018vsk}. It therefore also follows that in curved spacetime, this term is not irrelevant but can be absorbed  into a combination of the other covariant versions of the self-interactions $\mathcal{L}_{2,3,4,5}$ \cite{Jimenez:2016isa}.

To prove that \eqref{eqL6Tausgeschrieben} is indeed a total derivative (in four-dimensional flat spacetime), it is useful to first define the following tensor
\begin{equation}
X^\mu{}_\nu=\epsilon^{\mu\rho\sigma\chi}\epsilon^{\alpha\beta\delta\kappa}B_{\nu\alpha}B_{\rho\beta}B_{\sigma\delta}B_{\chi\kappa}\,.
\end{equation}
Then, for any arbitrary four-dimensional tensor $B\mn$, (not necessarily symmetric), the following non-trivial identity holds\footnote{This relation is trivial to show for symmetric tensors as they are diagonalizable, but it holds true as an identity even for non-symmetric tensors $B\mn$. Indeed, if we first focus on the diagonal, it should be clear that for any arbitrary $4\times 4 $ matrix $B_{ab}$ (not necessarily symmetric), one has 
\ba
X^0{}_0=X^1{}_1=X^2{}_2=X^3{}_3=3! \det B\,,
\ea
with $\det B=\frac{1}{4!}X^\mu{}_\mu$. Next consider an arbitrary off-diagonal element, for instance $X^0{}_1$, then
\ba
X^0{}_1&=&\epsilon^{0 b' c' d'}\epsilon^{a b c d}B_{1a}B_{b'b}B_{c'c}B_{d'd}\\
X^0{}_1&=&3! \epsilon^{a b c d}B_{1a}B_{1b}B_{2c}B_{3d}\equiv 0\,,
\ea
therefore all off-diagonal necessarily vanish from the antisymmetry property of the Levi-Cevita symbol, proving the  standard identity \eqref{identityRelation}.
}
\begin{equation}\label{identityRelation}
X^\mu{}_\nu=\frac14 X^\alpha{}_\alpha \delta^\mu{}_\nu\,.
\end{equation}
In terms of $X^\mu{}_\nu$, the variation of $\mathcal{L}_6^T$ with respect to $A_\mu$, is simply
\begin{align}
\frac{\delta  \mathcal{L}_6^T}{\delta A_\mu}&= \frac{\delta}{\delta A_\mu} \big[f_6(A^2)\epsilon^{\mu\nu\rho\sigma}\epsilon^{\alpha\beta\delta\kappa}\partial_\mu A_\alpha\partial_\nu A_\beta \partial_\rho A_\delta \partial_\sigma A_\kappa  \big]\nonumber\\
&=  -2\cdot4\, f_6' A^\gamma X^\mu{}_\gamma + 2f_6' A^\mu X^\alpha{}_\alpha \,,
\end{align}
where we commit again to $B\mn=\p_\mu A_\nu$.
Now, a crucial point is to use the identity relation \eqref{identityRelation},
leading to
\begin{align}
\frac{\delta  \mathcal{L}_6^T}{\delta A_\mu}
&=  -2\cdot4f_6'\, A^\gamma X^\mu{}_\gamma + 2f_6' A^\mu X^\alpha{}_\alpha \nonumber\\
&=  -2\cdot4f_6'\, A^\gamma \frac{X^\alpha{}_\alpha\delta^\mu_\gamma}{4} + 2f_6' A^\mu X^\alpha{}_\alpha  \nonumber\\
&\equiv 0\,,
\end{align}
for any function $f_6$.
This proves that the term $\mathcal{L}_6^T$ advocated as a genuinely new interaction is just a total derivative and does not give any non-trivial contribution. This is in agreement with the decoupling limit analysis presented in \cite{Jimenez:2016isa} and as argued in \cite{Heisenberg:2018vsk}.

The total derivative which gives rise to this term can be easily constructed as
\ba
\mathcal{L}_6 &\supset &\frac{1}{4!} \partial_\mu\Big(g_6(A^2) \epsilon^{\mu\nu\rho\sigma}\epsilon^{\alpha\beta\gamma\delta}A_\alpha \partial_\nu A_\beta \partial_\rho A_\gamma \partial_\sigma A_\delta\Big)\nonumber\\
&= & \frac{2}{4!}g_6'A^\lambda A_\alpha\epsilon^{\mu\nu\rho\sigma}\epsilon^{\alpha\beta\gamma\delta} B_{\mu \lambda} B_{\nu \beta} B_{\rho \gamma} B_{\sigma \delta}\nonumber\\
 & +& g_6\det B \nn \\
& = & \frac{1}{2}g_6' A^\lambda A_\alpha\Big(\frac{\partial \det B}{\partial B} B\Big)^\alpha{}_\lambda+g_6 \det B.
\ea
We can now use that\footnote{The derivative of the determinant allows to give another proof of \eqref{identityRelation} by noticing that
\begin{equation}
X=\frac{4!}{4}\frac{\partial\det B}{\partial B}B=\frac{4!}{4}\det B\,{\mathbbm{1}}=\frac14[X]\,{\mathbbm 1}.
\end{equation}} $\frac{\partial \det B}{\partial B}=\det B  B^{-1}$ to finally obtain
\begin{equation}
\frac{1}{4!}\partial_\mu\Big(g_6 \epsilon^{\mu\nu\rho\sigma}\epsilon^{\alpha\beta\gamma\delta}A_\alpha \partial_\nu A_\beta \partial_\rho A_\delta \partial_\sigma A_\kappa\Big)=\partial_Y(g_6Y)\det  B.
\end{equation}
that is $\mathcal{L}_6^T$ in \eqref{LagrangianL6ind} with $f_6=\partial_Y(g_6Y)$ and $Y=A^2$.
Thus, having shown that this term is a total derivative, this also permits to understand that the pure $S$ piece in $\mathcal{L}_6^T$, i.e. the $S^4$ order in \eqref{LagrangianL6ind}, can be written in terms of lower order interactions modulo a total derivative and, consequently, it does not contribute genuinely new interactions. This result gives further support to the findings in the decoupling limit performed in \cite{Jimenez:2016isa}, where it was shown that the would-be pure $S^4$ terms do not contribute new interactions in the decoupling limit, thus hinting that such interactions are in fact not present in the full theory. As argued there, at the fourth derivatives order, the decoupling limit only exhibits pure $F^4$ interactions and a mixing of the form $\tilde{F}\tilde{F}SS$, which arises from $\Lag_6^C$ above.

\section{Degenerate Hessian}
The presence of constraint equations is ensured by the degeneracy of the Hessian matrix defined by
\begin{equation}
\mathcal{H}^{\mu\nu}=\frac{\delta^2\mathcal{S}}{\delta\dot{A}_\mu\delta\dot{A}_\nu}\,,
\end{equation}
where its degeneracy means $\det \mathcal{H}^{\mu\nu}=0$. If this condition is fulfilled, it means that some of the degrees of freedom do not propagate and we actually have a constrained system. The crucial role of the Hessian in the counting of propagating degrees of freedom can be understood from different perspectives. In the Lagrangian formalism, the Hessian provides the principal part of the corresponding differential field equations so that having a degenerate Hessian translates into differential equations with a degenerate principal part. For the physical systems we have in mind (i.e., the propagation of some given fields that are to satisfy hyperbolic equations), the degeneracy of the principal part signals that we need fewer initial values to solve the differential equations because there is some non-hyperbolic sector that forces some constraints. In the Hamiltonian formulation, the Hessian determines the invertibility of the transformation from configuration space to phase space. Thus, a degenerate Hessian indicates again that the physical phase space is subject to a constrained surface that reduces the number of physical degrees of freedom.

However, the condition of having a degenerate Hessian by itself does not tell us how many constraints there are nor their nature. In order to elucidate that, one needs to obtain the full constraints algebra via the standard Dirac procedure \cite{Dirac}. In the language of Dirac's method, the degenerate Hessian indicates the presence of (at least one) primary constraints. Consistency of the equations requires the conservation of these primary constraints in the time evolution which will generate secondary constraints via their Poisson bracket with the Hamiltonian. Again, conservation of the secondary constraints can give rise to tertiary constraints and so on. After working out the full algebra of constraints, it may happen that some sub-algebra of constraints has a vanishing Poisson bracket with all the other constraints. In this case, such constraints are first class and they generate gauge symmetries so that they eliminate 4 phase space dofs (or 2 in configuration space) instead of 2 (1 in configuration space). The presence of first class constraints are at the heart of gauge theories and roots the difficulties of a covariant quantisation of such systems (see e.g \cite{Henneaux:1992ig}).

The importance of having constraints in theories with a vector field lies in the fact that the irreducible massive (massless) spin-1 representations of the Lorentz group has three (two) dofs, while an arbitrary vector field in four dimensions has four components. Furthermore, the would-be fourth polarisation has a ghostly nature since it can be associated to higher order field equations for the longitudinal mode. The simplest way of implementing a constraint on the temporal component of the vector field is by requiring $\mathcal{H}^{0\mu}=0$ that trivially gives the primary constraint enforcing that its conjugate momentum vanishes. In the most general case however there are some other non-trivial possibilities that could happen with the constraints:

\begin{itemize}

\item The required primary constraint generates a secondary constraint and that closes the constraints algebra with only second class constraints. In this case the vector field will propagate 3 polarisations. This is what happens for a massive Proca field.

 \item The primary constraint generates a secondary constraint, but we have a first class constraint. In this case the vector field features a gauge symmetry that is generated by the first class constraint. This is what happens for a Maxwell field and all the non-linear electrodynamic theories that are built in terms of the field strength. It is important to keep in mind however that this is not the only possibility and that the gauge symmetry could be realised in a wide variate of forms. 
  
 \item If we are interested in describing a spin-1 field, the above cases are the relevant ones. However, when studying a general Lagrangian written in terms of a vector field, the constraints algebra can reveal that there are fewer propagating degrees of freedom so that the theory can actually describe a field of a different nature that is related to $A_\mu$ via some non-trivial constraints or dualities.
  
  \end{itemize}
  
  In the following we illustrate the above points through various arguments and examples. 
  
\subsection{Field redefinitions and dualities}
  
  In order to illustrate the previous point, let us take a Maxwell theory with $\Lag_{\rm M}=-\frac14 F^2$ with the  $U(1)$ gauge symmetry $\delta_\theta A_\mu=\partial_\mu\theta$ and perform a field redefinition of the form $A_\mu\rightarrow A^2 A_\mu$. The Maxwell Lagrangian then becomes
\begin{align}\label{complicatedMaxwell}
\Lag_{\rm M}=&-\frac{A^2}{4}\Big(A^2\eta^{\mu\nu}+6 A^{\mu}A^\nu\Big)F_{\mu\lambda} F_\nu{}^\lambda\nonumber\\
&-\frac{1}{2}A^\mu A^\nu\Big(A^2\eta^{\alpha\beta}-A^\alpha A^\beta\Big)S_{\mu\alpha}S_{\nu\beta}\nonumber\\
&+2A^2A^\mu A^\nu F_{\mu\lambda} S_\nu{}^\lambda.
\end{align}
Although not trivial at all, this Lagrangian obviously features a gauge symmetry inherited from the original $U(1)$ symmetry whose infinitesimal realisation in the new field variable reads $\delta_\theta A_\mu=A^{-2}(\delta_\mu{}^\nu-\frac{2}{3A^2} A_\mu A^\nu)\partial_\nu\theta$. As a matter of fact, we can see interactions that might look problematic at first sight.
However, we know that this is just Maxwell's theory in disguise and, therefore, despite its worrisome look, it is perfectly healthy, it propagates 2 polarisations and it enjoys a gauge symmetry. An analogous finding can drawn after applying a Vector duality transformation to the Maxwell Lagrangian as shown in \cite{deRham:2014lqa}. Combining the fact that any non-linear electrodynamics theory has the $U(1)$ gauge symmetry together with a general field redefinition, it is straightforward to understand that there is a whole plethora of gauge-invariant theories where the gauge symmetry is not apparent. The Dirac method however systematically detects the presence of the gauge symmetry. An interesting question that arises here is whether the presence of an abelian one-parameter gauge symmetry must correspond to the usual $U(1)$ gauge symmetry via some field redefinition. This problem was considered in \cite{Wald:1986bj} where it was concluded that any non-linear deformation of the Maxwell Lagrangian with some Bianchi identities associated to a gauge symmetry, must in fact have the usual $U(1)$ gauge invariance, with the possibility of field redefinitions. A caveat of that analysis is that the potentially deformed gauge symmetry is restricted to depend up to first derivatives of the field and at most linearly. The problem of finding modified Maxwell Lagrangians with two or fewer degrees of freedom has also been recently considered in \cite{Mukohyama:2019unx}.

Similarly, theories describing a massive vector field will be prone to ambiguities due to field redefinitions so that before disregarding a given Lagrangian containing harmful looking individual interactions one should check for non-trivial relations rendering the full Lagrangian (perhaps up to a given scale) viable. For illustrative purposes, let us consider the Lagrangian\footnote{This is perhaps the simplest non-trivial example. Adding a mass (or a general potential) to the Maxwell Lagrangian does not bring anything new for our discussion other than breaking the gauge symmetry. The term $f_3(A^2) [S]$ remains the same (up to total derivatives) under the considered field redefinition.}
\begin{equation}
\Lag_4=A^2\Big([S^2]-[S]^2\Big)
\label{L4proxy}
\end{equation}
and, as before, perform the field redefinition $A_\mu\rightarrow A^2 A_\mu$ that transforms it into
\begin{align}
\frac{1}{A^{10}}\Lag_4=\Big(&\eta^{\mu\alpha}\eta^{\nu\beta}-\eta^{\mu\nu}\eta^{\alpha\beta}-4 \hA^\mu \hA^\nu\eta^{\alpha\beta}+6\hA^\mu \hA^\alpha \eta^{\nu\beta}\nonumber\\
&-2 \hA^\mu \hA^\nu \hA^\alpha \hA^\beta\Big)S_{\mu\nu} S_{\alpha\beta}+2\hA^\mu\hA^\nu F_{\mu\alpha} F_\nu{}^{\alpha}\nonumber\\
&-8\hA^\mu\hA^\nu F_{\mu\alpha} S_\nu{}^{\alpha},
\end{align}
where we have introduced the unit norm vector $\hA_\mu\equiv A_\mu/\sqrt{|A^2|}$. We again see the appearance of worrisome terms, but we can be confident that the different terms will conspire to guarantee the propagation of three polarisations since this is related to \eqref{L4proxy} via a field redefinition. The ambiguities due to field redefinitions can be evaded by looking at physical quantities such as scattering amplitudes. A key ingredient in many of those formulations is how the field ultimately couples to external sources and one should really fix those from the outset to avoid field redefinitions redundancies.

The above situations give the possibilities that eventually describe a (massive or massless) spin-1 field. These however do not exhaust all the possibilities for a general Lagrangian written in terms of a vector field. In addition to field redefinitions of the vector field, we can make use of dualities like e.g. the Hodge duality that relates a 1-form and a 3-form in four dimensions via $C= \star A$. It is well-known that a massless 3-form only propagates a global degree of freedom, while a massive 3-form propagates one degree of freedom. The Hodge dual formulation of a 3-form is in terms of a 1-form and, consequently, another situation that could arise when analysing a general action for a vector field is that it could simply be the dual formulation of some massive or massless 3-form field. In the massless case, the gauge symmetry of the 3-form $C\rightarrow C+\d\theta$, with $\theta$ an arbitrary 2-form, translates into a symmetry of the form $A\rightarrow A+\star\d\theta$ for the 1-form. The simplest example of this is the Lagrangian $\Lag=\frac12(\partial_\mu A^\mu)^2+\frac12 m^2 A^2$, which is the Hodge dual Lagrangian of a massive 3-form. In the massless case we see that the equation of motion implies $\partial_\mu A^\mu$ be constant and it has the gauge symmetry $A_\mu\rightarrow A_\mu +\epsilon_{\mu\alpha\beta\gamma} \partial^\alpha \theta^{\beta\gamma}$. In the massive case, the theory propagates one degree of freedom. This can be straightforwardly shown by introducing a \stu field $b^{\mu\nu}$ via the replacement $A_\mu=\tilde{A}_\mu +\frac{1}{m}\epsilon_{\mu\alpha\beta\gamma} \partial^\alpha b^{\beta\gamma}$ and taking the decoupling limit $m\rightarrow 0$ so that the Lagrangian contains a massless 3-form sector $\frac12(\partial_\mu \tilde{A}^\mu)^2$ and a massless 2-form sector $\partial_{[\mu}b_{\nu\rho]} \partial^{[\mu}b^{\nu\rho]}$ which, as it is well-known, can be dualised to a scalar field. Again, by performing field redefinitions of $A_\mu$, one can generate very contrived Lagrangians that nevertheless secretly describe a (massive or massless) 3-form field.

\subsection{Order by order or mixed Lagrangians}
In \cite{Jimenez:2016isa} the degeneracy of the Hessian and the construction of derivative self-interactions were studied order by order in powers of $S_{\alpha\beta}$ and the focus was given to avoid propagating $A_0$ or higher order derivatives for the longitudinal mode, i.e., the interactions were constructed so that the longitudinal mode in the decoupling limit was restricted to be within the Horndeski class. Of course, this restriction is not fundamental and can be easily dropped for instance by allowing the longitudinal mode to be in the wider class of degenerate scalar-tensor theories not contained in the Horndeski family. As a matter of fact, the condition to have a degenerate Hessian is non-linear and, therefore, it can be fulfilled even if it is not satisfied at each order in $S_{\mu\nu}$.
Similarly, the statement that the Horndeski Lagrangian is the most general Lagrangian for a scalar field has to be understood in a way that one studies the constraints algebra at each individual order in $\Pi_{\mu\nu}=\partial_\mu\partial_\nu\pi$ and performs in this way a more restricted construction of the allowed interactions.

The Lagrangian \eqref{generalizedProfaField} with the interactions \eqref{vecGalProcaField} is the general Lagrangian for a massive vector theory with at most three propagating degrees of freedom, if the degeneracy of the Hessian matrix is imposed at each individual order in $\mathcal L_i$. Furthermore, more general looking terms can be generated from the individual Lagrangians $\mathcal L_i$ if a disformal transformation is performed\footnote{This is specially relevant when including gravity where the disformal transformation corresponds to a field redefinition. A disformal transformation including a vector can actually be much more general than the one given here.} $\eta_{\mu\nu}\to\eta_{\mu\nu}+g(A^2)A_\mu A_\nu$. However, they do not represent any genuinely new interaction since one is performing a simple change of the background metric and they simply are connected to the original interactions via a disformal relation.

In \cite{deRham:2018svs} new interactions not contained in generalized Proca theories were constructed with the help of taking the AdS decoupling limit of massive gravity. The crucial difference between the consistency of these new interactions and the standard generalized Proca interactions is whether the degeneracy condition is satisfied order by order for each individual Lagrangians or by a combination of the Lagrangians, as it is the case for instance in \cite{deRham:2014lqa}.
Even if the individual terms $H_1$ and $H_2$ do not possess the necessary primary constraint to render the temporal component of the vector field non-dynamical, the combination of them $H_1+H_2$ can allow it in some cases through very non-trivial cancellations between the various orders. This was already pointed out in \cite{Kimura:2016rzw} even in the general case of curved space-time. Specially, the presence of $A^\mu A^\nu F_{\mu\lambda} S_\nu{}^\lambda$ type of interactions, which would be problematic by its own, can be rendered harmless in the simultaneous presence of other interactions among different orders (see for instance \cite{deRham:2010gu}). As we saw in equation \eqref{complicatedMaxwell}, this class of interactions can be related to standard Maxwell by means of field redefinition $A_\mu\rightarrow A^2 A_\mu$. As shown in \cite{deRham:2018svs}, there is however a set of interactions that cannot be directly related to  Generalized Proca via field redefinitions of the form $A_\mu\rightarrow A^2 A_\mu$ nor via a Vector duality as in \cite{deRham:2014lqa} but they are still very likely related to the standard generalized Proca theories by means of another type of duality relation which is being studied somewhere else.

In order to illustrate the point made above, let us consider a theory for a vector field whose Lagrangian can be written as a series in derivatives of the field
\begin{equation}
\Lag=\sum_{n\geq0}\mM_{n}^{\mu_1\nu_1\cdots\mu_n\nu_n} \partial_{\mu_1}A_{\nu_1}\cdots \partial_{\mu_n}A_{\nu_n}
\end{equation}
where $\mM_n^{\mu_1\nu_1\cdots\mu_n\nu_n}$ is some tensor that can depend on $A_\mu$, but not its derivatives and the sum can extend to infinity. The existence of constraints as well as their nature will thus be determined by the properties of $\mM_n^{\mu_1\nu_1\cdots\mu_n\nu_n}$. The Hessian can then be straightforwardly computed
\begin{equation}
\mathcal{H}^{\mu\nu}=\sum_{n\geq2}n(n-1)\mathcal{M}_n^{0\mu0\nu\mu_3\nu_3\cdots\mu_n\nu_n} \partial_{\mu_3}A_{\nu_3}\cdots \partial_{\mu_n}A_{\nu_n}.
\end{equation}
Notice that neither $\mM_0$ nor $\mM_1^{\mu\nu}$ contribute to the Hessian so that the degeneracy condition does not depend on them. This simply reflects that $\mM_0$ is a potential term while $\mM_1^{\mu\nu}$ describes the term $f_3(A^2)\partial_\mu A^\mu$ so they do not contribute to the propagating sector. They can however crucially affect the nature of the existing constraints. We will write the Hessian in the more compact form
\begin{equation}
\mathcal{H}^{\mu\nu}=\sum_{n\geq0}\mathcal{H}^{\mu\nu}_n,
\end{equation}
with an obvious definition of $\mH_n$ in terms of $\mM_{n-2}$. Its determinant is then
\begin{equation}
4!\det \mH^{\mu\nu}=\epsilon_{\alpha_1\cdots\alpha_4}\epsilon_{\beta_1\cdots\beta_4}\sum_{m,n,p,q}\mH^{\alpha_1\beta_1}_m\mH^{\alpha_2\beta_2}_n\mH^{\alpha_3\beta_3}_p\mH^{\alpha_4\beta_4}_q.
\end{equation}
This expression will suffice for us since we do not intend to give an exhaustive analysis, but rather we want to explicitly show the possibility of a non-trivial interplay between the different orders that give a degenerate Hessian. Let us write the first terms of the expansion explicitly to make it more clear
\begin{align}
4!\det \mH^{\mu\nu}=&\;\epsilon_{\alpha_1\cdots\alpha_4}\epsilon_{\beta_1\cdots\beta_4}
\Big[\mH^{\alpha_1\beta_1}_0\mH^{\alpha_2\beta_2}_0\mH^{\alpha_3\beta_3}_0\mH^{\alpha_4\beta_4}_0\nonumber\\
&+4\mH^{\alpha_1\beta_1}_0\mH^{\alpha_2\beta_2}_0\mH^{\alpha_3\beta_3}_0\mH^{\alpha_4\beta_4}_1\nonumber\\
&+4\mH^{\alpha_1\beta_1}_0\mH^{\alpha_2\beta_2}_0\mH^{\alpha_3\beta_3}_0\mH^{\alpha_4\beta_4}_2\nonumber\\
&+6\mH^{\alpha_1\beta_1}_0\mH^{\alpha_2\beta_2}_0\mH^{\alpha_3\beta_3}_1\mH^{\alpha_4\beta_4}_1\nonumber\\
&+4\mH^{\alpha_1\beta_1}_0\mH^{\alpha_2\beta_2}_0\mH^{\alpha_3\beta_3}_0\mH^{\alpha_4\beta_4}_3\nonumber\\
&+12\mH^{\alpha_1\beta_1}_0\mH^{\alpha_2\beta_2}_0\mH^{\alpha_3\beta_3}_1\mH^{\alpha_4\beta_4}_2\nonumber\\
&+4\mH^{\alpha_1\beta_1}_0\mH^{\alpha_2\beta_2}_1\mH^{\alpha_3\beta_3}_1\mH^{\alpha_4\beta_4}_1+\cdots\Big]
\label{degcondpert}
\end{align}
that clearly shows how the degeneracy of the Hessian can be achieved by non-trivial cancellations between terms of different orders in derivatives. Thus, we can compute the degeneracy condition order by order in the derivative expansion. At the lowest order\footnote{We refer to the order in derivatives in the Hessian, which corresponds to two higher orders in derivatives in the Lagrangian. Thus, the zeroth order in the Hessian corresponds to $(\partial A)^2$ in the Lagrangian and so on.} we will need to impose the vanishing of $\det\mH_0^{\mu\nu}$. If $\mH_0^{0\mu}=0$ the determinant trivially vanishes, but we can have a less trivial cancellation as it occurs for the Lagrangians given in \eqref{complicatedMaxwell} and \eqref{L4proxy}. At the next order, we will need to cancel the contribution $\sim\mH_0^3\mH_1$ and this will be enough to have the constraint at that order. However, this might not guarantee the persistence of this constraint in the full theory because the first order will also contribute to the quadratic order in the degeneracy condition (fourth line in the above expression). We will then need to construct the second order contribution determining $\mH_2$ so that the constraint is maintained at that order as well, i.e., such that the third and fourth lines in \eqref{degcondpert} cancel among them. Again, the constraint can only be guaranteed at second order, since $\mH_2$ will also contribute at cubic order so that $\mH_3$ will have to be such that it cancels that contribution. In principle, one can systematically use this perturbative procedure to guarantee the existence of the constraint at some given order.

\section{Discussion}
In this note we have reconsidered the derivative interactions of a single massive Proca field in flat spacetime with the aim of clarifying some potentially confusing points. We have explicitly shown that the pure $S^4$ terms do not  give rise to new interactions by showing the total derivative nature of $\Lag_6^T=f_6(A^2)\epsilon^{\mu\nu\rho\sigma}\epsilon^{\alpha\beta\delta\kappa}\partial_\mu A_\alpha\partial_\nu A_\beta \partial_\rho A_\delta \partial_\sigma A_\kappa$. We have then discussed the degenerate condition of the Hessian and how the structure and nature of the constraints guarantee the propagation of a massless or massive spin-1 field. The role of field redefinitions and dualities has also been considered. In particular, how the gauge symmetry in the massless case can be non-trivially realised or how the vector field can actually describe fewer than two degrees of freedom corresponding to a massless or massive three form via a duality. Finally, we have explained how the Hessian can be degenerate by non-trivial cancellations among different orders in the Lagrangian even if they do not individually produce a degenerate Hessian. This has been explicitly illustrated with a perturbative scheme that clearly shows the non-trivial interplay between different orders.

\acknowledgments JBJ acknowledges support from the  {\it Atracci\'on del Talento Cient\'ifico en Salamanca} programme and the MINECO's projects FIS2014-52837-P and FIS2016-78859-P (AEI/FEDER). CdR thanks the Royal Society for support at ICL through a Wolfson Research Merit Award. CdR is supported by the European Union’s Horizon 2020 Research Council grant 724659 MassiveCosmo ERC-2016-COG and by a Simons Foundation award ID 555326 under the Simons Foundation’s Origins of the Universe initiative, ‘Cosmology Beyond Einstein’s Theory’.
LH is supported by funding from the European Research Council (ERC) under the European Unions Horizon 2020 research and innovation programme grant agreement No 801781 and by the Swiss National Science Foundation grant 179740.


\end{document}